\documentclass[journal,article,submit,pdftex,moreauthors]{Definitions/mdpi} 

\usepackage{placeins}  % \FloatBarrier
\usepackage{subfig}

\firstpage{1} 
\pubvolume{1}
\issuenum{1}
\articlenumber{0}
\pubyear{2025}
\copyrightyear{2025}
\datereceived{ } 
\daterevised{ } % Comment out if no revised date
\dateaccepted{ } 
\datepublished{ } 
\hreflink{https://doi.org/} % If needed use \linebreak
\Title{Why Some Strikes Last Longer: The Resilience Paradox of Partially Functional Institutions}

\TitleCitation{Title}

\Author{Nuno Crokidakis $^{1,}$*\orcidA{}}

\AuthorNames{Firstname Lastname, Firstname Lastname,  and Firstname Lastname}
\AuthorCitation{Lastname, F.; Lastname, F.; Lastname, F.}
\address{%
$^{1}$ \quad Instituto de F\'isica, Universidade Federal Fluminense, Niter\'oi/RJ, Brazil
}

\corres{Correspondence: nunocrokidakis@id.uff.br}%; Tel.: (optional; include country code; if there are multiple corresponding authors, add author initials) +xx-xxxx-xxx-xxxx (F.L.)}

\abstract{
Labor strikes do not necessarily lead to complete institutional disruption, particularly in organizations where non-striking sectors can maintain a substantial fraction of normal activities. We propose a minimal dynamical model to investigate how such residual activity affects the persistence of labor conflicts, motivated by strikes in public universities where administrative staff may suspend their activities while academic personnel continue working. The model couples strike participation, institutional response, and overall functionality, incorporating two competing mechanisms: a direct suppressive effect of continued activity on strike participation and an indirect effect arising from the reduced institutional pressure to resolve the conflict. Analytical calculations reveal a counterintuitive regime in which increasing the activity of the non-striking sector increases, rather than decreases, stationary strike participation. We characterize this behavior through a strike-response susceptibility and derive an exact compensation threshold separating the conventional regime, in which residual activity suppresses strike participation, from a resilience-paradox regime, in which the opposite response occurs. At the compensation threshold, the direct and indirect mechanisms exactly balance, making stationary strike participation independent of the activity maintained by the non-striking sector. Numerical results confirm the analytical predictions and illustrate the reversal of the collective response across the threshold. These results show that institutional resilience can have an unintended dynamical consequence: by absorbing part of the disruption generated by a strike, continued operation may reduce the urgency of institutional response and thereby favor conflict persistence. The proposed mechanism provides a simple example of how individually stabilizing processes can collectively generate counterintuitive outcomes in complex social systems.
}

\vspace{1cm} 
\keyword{Labor strikes, Institutional resilience, Resilience paradox, Nonlinear dynamics, Collective behavior, Complex social systems}

\begin{document}

%%%%%%%%%%%%%%%%%%%%%%%%%%%%%%%%%%%%%%%%%%
%\setcounter{section}{-1} %% Remove this when starting to work on the template.

\tableofcontents

\section{Introduction}

Labor strikes constitute one of the main collective mechanisms through which workers can exert pressure during labor disputes. By withholding labor, workers impose economic, organizational, or social costs on employers and institutions, thereby altering the balance of bargaining power. For this reason, the occurrence and duration of strikes have long attracted attention in labor economics and industrial relations. Classical approaches have interpreted strike activity in terms of bargaining processes, conflicting expectations, asymmetric information, and the costs incurred by the parties while the dispute persists \cite{Ashenfelter1969,Kennan1985,Kennan1986,Tracy1987,Cramton1992}. In particular, the duration of a strike is closely related to the capacity of each side to absorb the costs generated by the conflict and to maintain its bargaining position over time.

A central element underlying these approaches is the disruptive character of a strike. The withdrawal of labor reduces production or service provision, generating pressure for a settlement. Empirical and theoretical studies have consequently emphasized the role of strike costs, bargaining power, replacement strategies, and the ability of the parties to withstand prolonged disputes \cite{CardOlson1995,SchnellGramm1994}. More recent work on worker power similarly stresses that the ability to generate disruption is one of the principal sources of leverage available to workers during collective action \cite{Kallas2026}. In public-sector disputes, however, the connection between work stoppage and institutional disruption can be particularly complex, since the continuation of essential or complementary activities may preserve part of the institution's normal functioning even during a strike \cite{Cordova1985,Rose2008}.

This situation naturally connects labor disputes with the broader concept of organizational resilience. Resilience is commonly associated with the ability of a system or organization to preserve functionality when exposed to disturbances or adverse conditions \cite{Holling1973,LengnickHall2011}. In organizational settings, resilience is generally regarded as a desirable property because it allows institutions to continue operating despite external or internal disruptions. Such a perspective has been developed in several areas, where resilience is understood in terms of the capacity to maintain essential functions, adapt to shocks, and recover from disturbances.

In the context of labor conflicts, however, resilience may have a less obvious consequence. If part of an institution continues to operate during a strike, the immediate disruption produced by the striking sector is partially absorbed. Although this residual activity helps preserve institutional functionality, it may simultaneously reduce the urgency of the institutional response. The resulting feedback raises a simple but largely unexplored question: can greater institutional resilience contribute to the persistence of a strike? In other words, a mechanism that is locally stabilizing from the perspective of institutional functioning may have the opposite effect on the dynamics of the underlying conflict.

Questions of this type are particularly suited to a dynamical-systems perspective. Statistical-physics and complex-systems approaches have shown that simple interactions among social components can generate collective outcomes that are not immediately inferable from the behavior of the individual mechanisms involved \cite{Castellano2009,Perc2017,nuno1,nuno2,nuno3}. Minimal models are especially useful in this context because they allow competing feedbacks to be isolated and their macroscopic consequences to be analyzed explicitly.

Motivated by this idea, we introduce a minimal dynamical model for labor strikes in partially functional institutions. The model considers a striking sector, a non-striking sector that maintains a fraction of institutional activity, and an institutional response driven by the loss of functionality. Residual activity affects strike participation through two competing pathways: it directly suppresses strike participation, but it also preserves institutional functionality and thereby weakens the institutional response to the strike. We show analytically that the competition between these mechanisms gives rise to two qualitatively distinct regimes separated by an exact compensation threshold. In the conventional regime, increasing residual activity reduces stationary strike participation. In contrast, when the indirect resilience-mediated feedback dominates, increasing residual activity enhances stationary strike participation. We refer to this counterintuitive behavior as the \textit{resilience paradox}. A strike-response susceptibility is introduced to characterize the sign and intensity of this response, and numerical results illustrate the mechanism underlying the two regimes.

%Long labor conflicts are often interpreted as a consequence of negotiation failures or irreconcilable demands. However, an alternative explanation is possible. In partially functional institutions, the continuation of essential activities may reduce the urgency for conflict resolution, creating conditions for the persistence of intermediate states in which labor mobilization and institutional operation coexist.

%%%%%%%%%%%%%%%%%%%%%%%%%%%%%%%%%%%%%%%%%%

\section{The Model}

We consider a partially functional institution undergoing a labor conflict. The main motivation comes from strikes in public universities, where a fraction of the administrative staff may interrupt their activities while other sectors, such as the academic staff, continue working. As a consequence, the institution does not necessarily become fully paralyzed: teaching, research, and part of the administrative activity may coexist with an ongoing strike. Our aim is not to describe the emergence of labor grievances or the decision to initiate a strike, but rather to investigate the dynamical mechanisms that may control its persistence once mobilization is already established.

Let $S(t) \in [0,1]$ denote the fraction of administrative workers participating in the strike. Accordingly, $1-S$ represents the fraction of this sector that remains active. We introduce a parameter $P \in [0,1]$ describing the activity maintained by a second sector that does not participate directly in the strike. In the university context motivating the model, $P$ may be interpreted as the level of academic activity maintained by faculty members. Importantly, $P$ does not represent political support for or opposition to the strike; it measures only the contribution of the non-striking sector to the continued operation of the institution.

The overall institutional functionality is represented by the macroscopic variable $U(t)$, defined as
\begin{equation} \label{eq1}
U(t) = \omega\,P + (1-\omega)[1-S(t)] ~,
\end{equation}
\noindent
where $0 \leq \omega \leq 1$ controls the relative contribution of the two sectors to institutional operation. Thus, $U=1$ corresponds to full functionality, whereas decreasing values of $U$  indicate increasing operational disruption. This definition captures a central feature of partially functional institutions: even for large values of $S$, the system may retain substantial functionality if the activity $P$ of the non-striking sector remains sufficiently high.

We assume that strike participation evolves according to
\begin{equation} \label{eq2}
\frac{dS}{dt} = \lambda(1-S) - \mu\,A\,S - \nu\,P\,S ~.
\end{equation}

The first term describes effective recruitment into the strike. Since the present model focuses on strike persistence rather than on the formation of the underlying grievances, the background level of worker dissatisfaction is absorbed into the effective mobilization rate $\lambda>0$. The factor $1-S$ ensures that recruitment is restricted to workers who have not yet joined the strike.

The second term represents the effect of institutional response. The variable $A(t) \geq 0$ quantifies the intensity of the institutional reaction to the conflict, including negotiation efforts, concessions, administrative actions, or other mechanisms capable of reducing strike participation. The parameter $\mu>0$ measures the effectiveness of this response. Finally, the term $\nu\,P\,S$ accounts for the direct effect of continued activity in the non-striking sector on strike participation, with $\nu>0$ measuring its strength. This term represents the possibility that continued institutional operation directly inhibits the maintenance or expansion of the strike.

The institutional response itself is assumed to be driven by the loss of functionality,
\begin{equation} \label{eq3}
\frac{dA}{dt} = \eta(1-U) - \theta\,A ~,
\end{equation}
\noindent
where $\eta>0$ measures the sensitivity of the institution to operational disruption and $\theta>0$ is the relaxation rate of the institutional response. Equation \eqref{eq3} embodies a key assumption of the model: the urgency to resolve the conflict increases when institutional functionality deteriorates. Conversely, an institution that remains substantially operational experiences a weaker pressure to intensify its response.

The complete dynamical system can thus be written as
\begin{eqnarray} \label{eq5}
\dot{S} & = & \lambda(1-S) - \mu\,A\,S - \nu\,P\,S \\ \label{eq6}
\dot{A} & = & \eta[(1-\omega)\,S + \omega\,(1-P)] - \theta\,A  
\end{eqnarray}
\noindent
where the expression for $U(t)$, Eq. \eqref{eq1}, was introduced in Eq. \eqref{eq3}. The model contains two competing mechanisms associated with the continued activity $P$. First, $P$ acts directly against strike participation through the term $\nu\,P\,S$. On the other hand, increasing $P$ preserves institutional functionality, thereby reducing $1-U$ and consequently weakening the institutional response $A$. Since $A$ itself acts to reduce $A$, this generates an indirect feedback with the opposite effect. Thus, continued operation can simultaneously contribute to the suppression of strike participation and reduce the institutional urgency to resolve the conflict.

This competition is the central mechanism investigated in the following sections. In particular, it raises the possibility that institutional resilience, although beneficial in maintaining short-term functionality, may under appropriate conditions favor the persistence of the labor conflict. We refer to this counterintuitive mechanism as the \textit{resilience paradox}.

%%%%%%%%%%%%%%%%%%%%%%%%%%%%%%%%%%%%%%%%%%

\section{Results}

Let us consider the stationary states of the model, defined for $t\to\infty$. Starting from Eq. \eqref{eq1}, we have $U^* = \omega\,P + (1-\omega)(1-S^*)$. Considering this result in Eq. \eqref{eq6}, one obtains the stationary value
\begin{equation} \label{eq8}
A^* = \frac{\eta}{\theta}[(1-\omega)S^* + \omega(1-P)] 
\end{equation}

From Eq. \eqref{eq5}, the long-time limit gives $\lambda(1-S^*) - \mu\,A^*\,S^* -\nu\,P\,S^*$. Substituting Eq. \eqref{eq8} in this last results gives us a second-order polynomial for $S^*$, namely $a\,(S^*)^{2} + b\,S^* + c=0$. Defining $K=\mu\eta/\theta$, the parameters $a, b$ and $c$ can be written as
\begin{eqnarray} \label{eq10}
a & = & K(1-\omega) \\ \label{eq11}
b & = & \lambda + K\omega + (\nu-K\omega)\,P \\ \label{eq12}
c & = & -\lambda
\end{eqnarray}
From these results, one obtains
\begin{equation} \label{eq13}
S^*(P) = \frac{-b(P) + [b(P)^{2} + 4K(1-\omega)\lambda]^{1/2}}{2K(1-\omega)} ~,
\end{equation}
\noindent
where $b(P)$ is given by Eq. \eqref{eq11}.

To quantify how stationary strike participation $S^*$ responds to changes in the activity $P$ maintained by the non-striking sector, we define the strike-response susceptibility
\begin{equation} \label{eq14}
\chi_P= \frac{dS^*}{dP} ~.
\end{equation}
Positive values of $\chi_P$ indicate that increasing residual institutional activity increases stationary strike participation, whereas $\chi_P<0$ corresponds to the conventional response, in which continued activity suppresses strike participation. Performing the derivative defined in Eq. \eqref{eq14}, 
\begin{equation}  \label{eq15}
\chi_P = \frac{(\nu_c - \nu)\,S^*}{\sqrt{b(P)^{2} + 4K(1-\omega)\lambda}} ~,
\end{equation}  
\noindent
where we define the compensationl threshold
\begin{equation}  \label{eq16}
\nu_c = \frac{\mu\eta\omega}{\theta} ~.
\end{equation}  
\noindent
It defines a compensation threshold at which the direct suppressive effect of continued activity exactly balances the indirect resilience-mediated feedback.

The origin of the change in sign of $\chi_P$ can be understood by examining the two distinct pathways through which the residual activity $P$ affects stationary strike participation. The first pathway is direct. From Eq. \eqref{eq5}, continued activity contributes the term $-\nu\,P\,S$ to the dynamics of $S$. Therefore, when considered in isolation, an increase in $P$ enhances the suppression of strike participation. This is the conventional mechanism: a more operational institution provides less favorable conditions for the maintenance of the strike, leading to a decrease in $S^*$.

However, $P$ also affects strike participation through an indirect pathway mediated by institutional functionality and response. From Eq. \eqref{eq1}, increasing $P$ increases the overall functionality $U$. Since the institutional response is driven by the loss of functionality, $\dot{A} = \eta(1-U) - \theta\,A$, a larger $P$ reduces the stationary response $A^*$. In turn, because $A$ suppresses strike participation through the term $-\mu\,A\,S$, the reduction of $A^*$ weakens one of the mechanisms responsible for decreasing $S$. Thus, the indirect pathway has the opposite sign: increasing $P$ may ultimately increase $S^*$.

The two mechanisms can therefore be summarized schematically as:
\begin{equation}
P\uparrow
\quad\Longrightarrow\quad
\begin{cases}
-\nu P S \ \text{increases in magnitude}
& \Longrightarrow S^*\downarrow, \\[2mm]
U\uparrow \Longrightarrow A^*\downarrow
& \Longrightarrow S^*\uparrow.
\end{cases}
\end{equation}

The stationary behavior is determined by the competition between these two effects. Remarkably, the model allows this competition to be quantified exactly. The direct suppressive mechanism is controlled by $\nu$, whereas the effective strength of the indirect resilience-mediated mechanism is $\mu\eta\omega/\theta$. Their exact balance defines the compensation threshold $\nu_c$, given by Eq. \eqref{eq16}.

Consequently, the strike-response susceptibility $\chi_P$, changes sign at $\nu=\nu_c$. For $\nu>\nu_c$ the direct mechanism dominates ($\chi_P<0$), so that increasing residual activity reduces stationary strike participation, as conventionally expected. At $\nu=\nu_c$, the two effects compensate exactly ($\chi_P=0$), making $S^*$ independent of $P$ \footnote{This can also be seen in Eq. \eqref{eq11}. For $\nu=\nu_c, b(P)=b=\lambda+K\omega$, leading to $S^*=$ constant in Eq. \eqref{eq13}.}. In contrast, for $\nu<\nu_c$, the resilience-mediated feedback dominates and $\chi_P>0$.

In this regime, increasing the activity of the non-striking sector increases, rather than decreases, stationary strike participation. This counterintuitive response constitutes the resilience paradox predicted by the model.

Importantly, the paradox does not require any individual actor to behave paradoxically or to intentionally favor strike persistence. Continued activity locally contributes to institutional operation, while the institutional response rationally decreases when operational disruption becomes less severe. Nevertheless, the coupling between these individually reasonable mechanisms produces an unexpected collective outcome: by absorbing part of the disruption generated by the strike, institutional resilience reduces the pressure for a strong response and can thereby favor the persistence of the conflict. The paradox is therefore an emergent property of the feedback structure of the system rather than an assumption imposed on individual behavior.

%%%%%%%%%%%%%%%%%%%%%%%%%%%%%%%%%%%%%%%%%%%%%%%%%%%%%
\begin{figure}[t]
\centering
\vspace{0.5cm}
\includegraphics[width=0.6\linewidth]{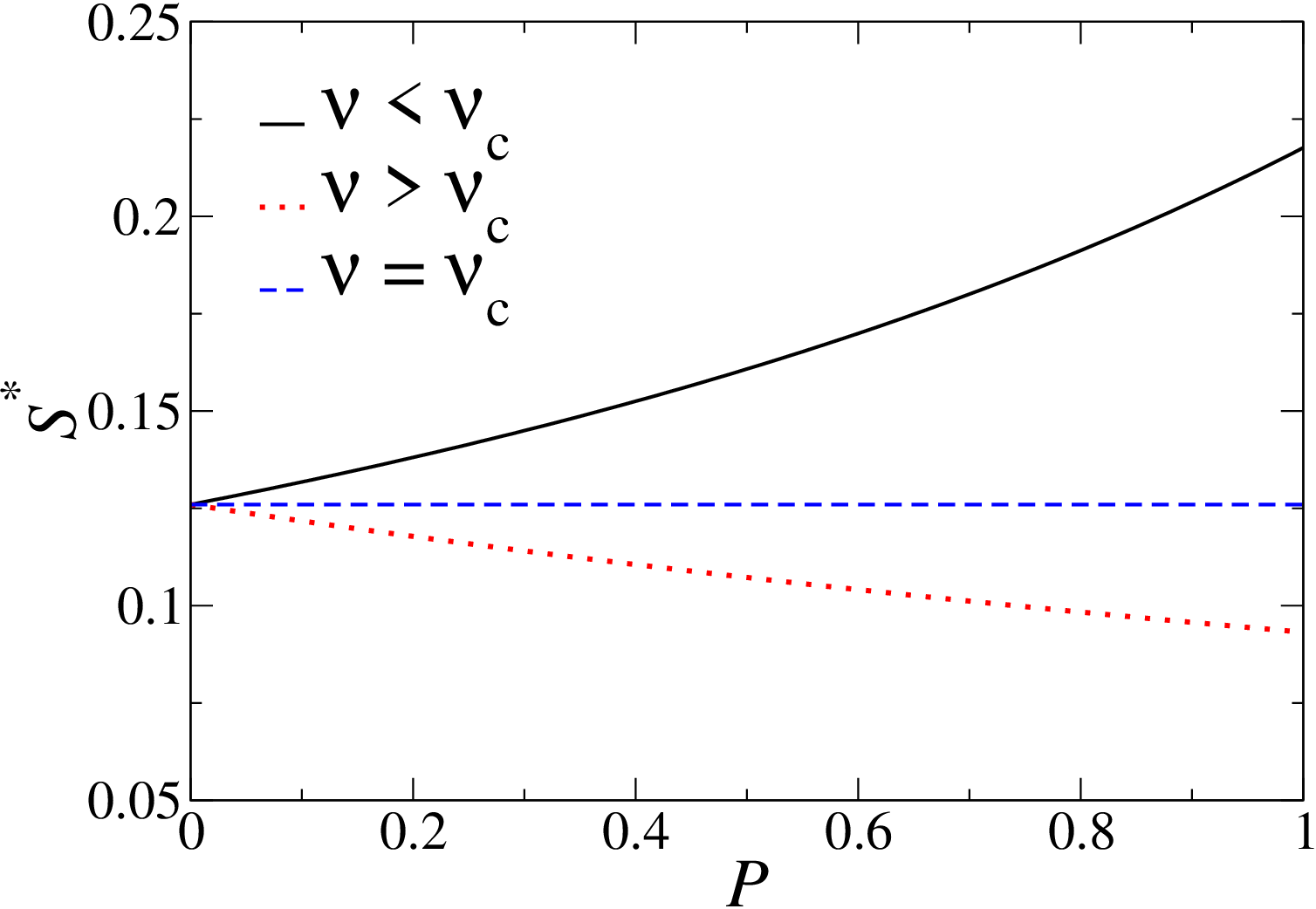}
\caption{Stationary strike participation $S^*$ as a function of the residual activity $P$ for different values of the direct suppression parameter $\nu$. The remaining parameters are $\lambda = 0.15, \mu = 0.60, \eta = 0.80, \theta = 0.30$ and $\omega = 0.60$, yielding the compensation threshold $\nu_c=0.96$, from Eq. \eqref{eq16}. For $\nu=0.40<\nu_c, S^*$ increases with $P$, corresponding to the resilience-paradox regime. At $\nu=\nu_c=0.96$, the direct and indirect mechanisms exactly compensate and $S^*$ becomes independent of $P$. For $\nu=1.40>\nu_c$, the conventional behavior is recovered and $S^*$ decreases with increasing $P$.}
\label{fig1}
\end{figure}
%%%%%%%%%%%%%%%%%%%%%%%%%%%%%%%%%%%%%%%%%%%%%%%%%%%%%

To verify the analytical results and to better understand the resilience paradox, we performed numerical calculations. In Fig. \ref{fig1}, we show the stationary strike participation $S^*$ as a function of the residual activity $P$ for different values of the direct suppression parameter $\nu$. The fixed parameters are $\lambda = 0.15, \mu = 0.60, \eta = 0.80, \theta = 0.30$ and $\omega = 0.60$, yielding the compensation threshold $\nu_c=0.96$ from Eq. \eqref{eq16}. For $\nu=0.40<\nu_c, \,S^*$ increases with $P$, corresponding to the resilience-paradox regime, as predicted analytically by Eq. \eqref{eq15}. At $\nu=0.96=\nu_c$, the direct and indirect mechanisms exactly compensate, and $S^*$ becomes independent of $P$. This result is also consistent with Eq. \eqref{eq15}, since $\chi_P=0$ at $\nu=\nu_c$. Finally, for $\nu=1.40>\nu_c$, where Eq. \eqref{eq15} predicts $\chi_P<0$, the conventional behavior is recovered, and $S^*$ decreases with increasing $P$.

\begin{figure}[t]
\centering
\vspace{0.5cm}
\includegraphics[width=0.6\linewidth]{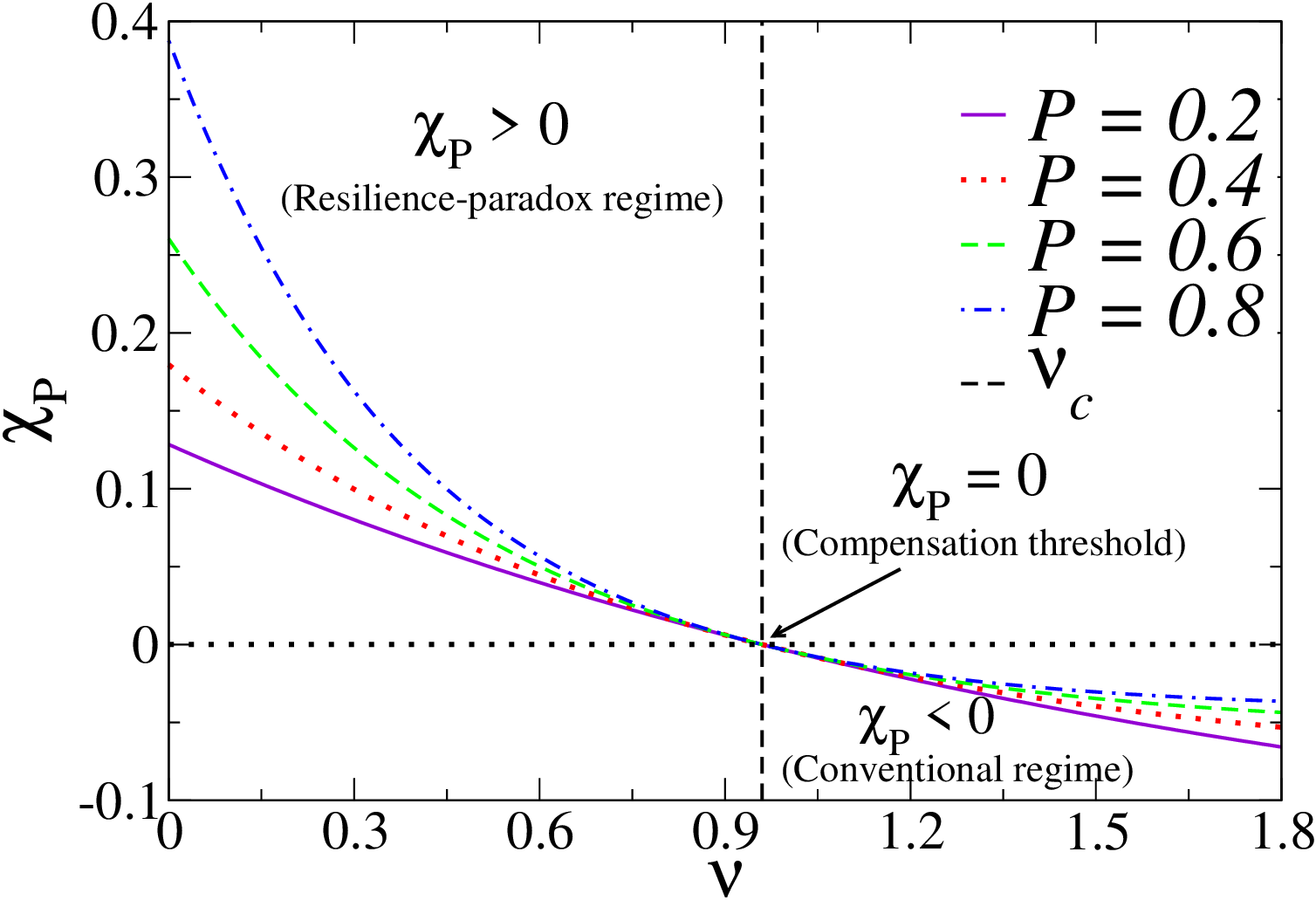}
\caption{Strike-response susceptibility $\chi_P$ as a function of the direct suppression parameter $\nu$, for different values of $P$. The remaining parameters are $\lambda = 0.15, \mu = 0.60, \eta = 0.80, \theta = 0.30$ and $\omega = 0.60$. The vertical dashed line indicates the compensation threshold $\nu_c=0.96$, given by Eq. \eqref{eq16}, at which the direct suppressive effect of residual activity exactly balances the indirect resilience-mediated feedback. All curves cross $\chi_P=0$ at the same threshold, independently of $P$. For $\nu<\nu_c, \chi_P>0$, corresponding to the resilience-paradox regime, in which increasing residual activity increases stationary strike participation. For $\nu>\nu_c, \chi_P<0$, and the conventional response is recovered, with increasing residual activity reducing stationary strike participation.}
\label{fig2}
\end{figure}

To characterize these regimes more direclty, we plot the strike-response susceptibility $\chi_P$ as a function of $\nu$ for different values of the residual activity $P$. The numerical results are exhibited in Fig. \ref{fig2}. The vertical dashed line marks the compensation threshold $\nu_c=0.96$. For $\nu<\nu_c$, we obtain $\chi_P>0$, corresponding to the resilience-paradox regime, in which increasing residual activity increases stationary strike participation. For $\nu>\nu_c$, $\chi_P<0$, and the conventional regime is recovered, with $S^*$ decreasing as $P$ increases. At $\nu=\nu_c$, all curves cross $\chi_P=0$, independently of $P$, in agreement with Fig. \ref{fig1} and Eq. \eqref{eq15}.

%%%%%%%%%%%%%%%%%%%%%%%%%%%%%%%%%%%%%%%%%%%%%%%%%%%%%
%%%%%%%%%%%%%%%%%%%%%%%%%%%%%%%%%%%%%%%%%%%%%%%%%%%%%

To gain further insight into the mechanism underlying the resilience paradox, we define two distinct suppression contributions,
\begin{eqnarray} \label{eq17}
F_{dir}(P) & = &  \nu\,P\,S^*  \\ \label{eq18}
F_{resp}(P) & = & \mu\,A^*\,S^*  
\end{eqnarray}

In terms of these quantities, Eq. \eqref{eq2} can be written at stationarity as
\begin{equation} \label{eq19}
0=\lambda(1-S^*) - F_{resp} - F_{dir} 
\end{equation}
The quantities $F_{dir}(P)$ and $F_{resp}(P)$ therefore represent the direct and response-mediated contributions to the suppression of strike participation, respectively. Since both depend of $P$, they provide a useful way to visualize the competition between the two mechanisms responsible for the collective response of the system.

\begin{figure}[t]
\begin{center}
\vspace{6mm}
\includegraphics[width=0.45\textwidth,angle=0]{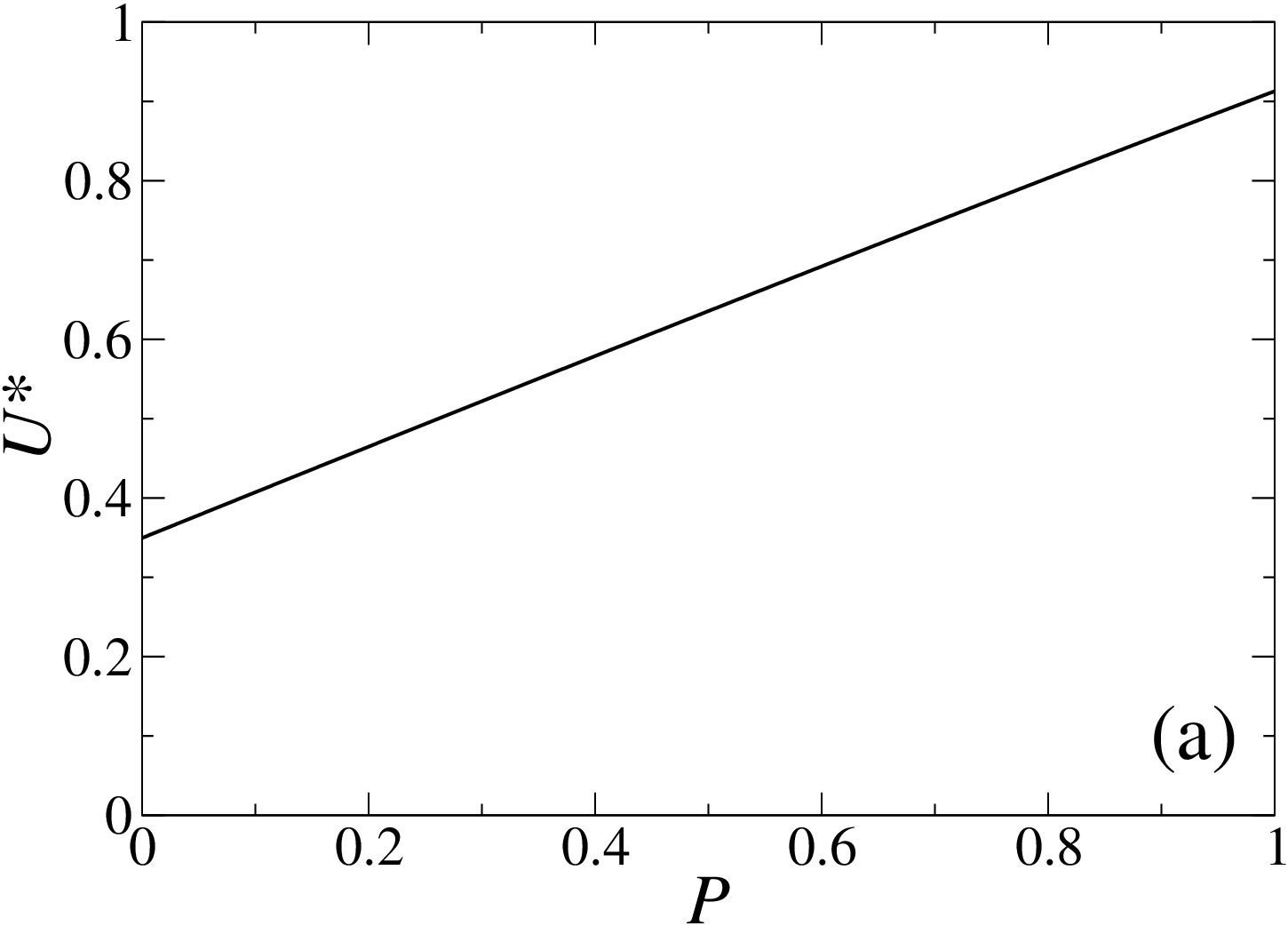}
\hspace{0.3cm}
\includegraphics[width=0.45\textwidth,angle=0]{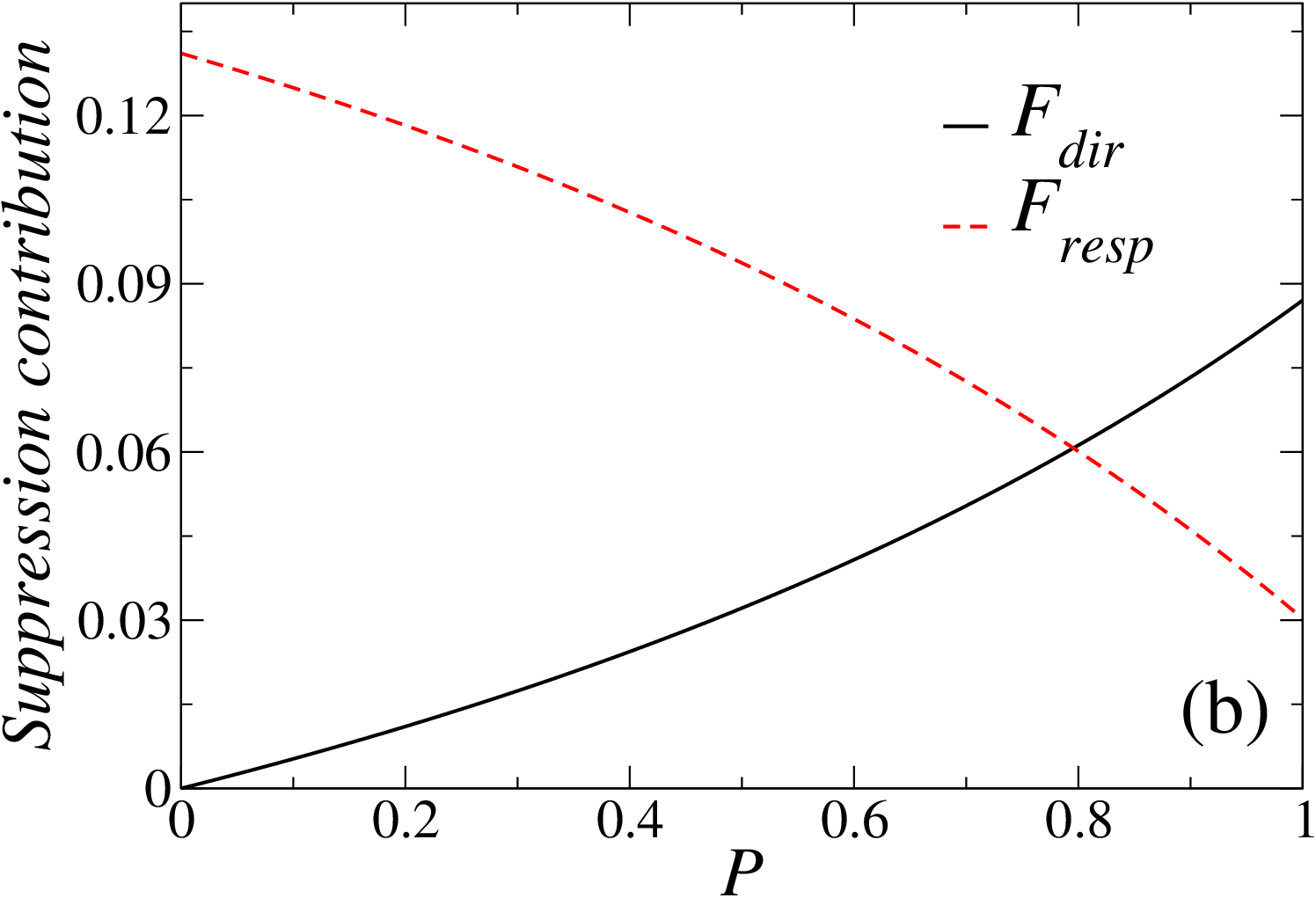}
\end{center}
\caption{Mechanism underlying the resilience paradox. (a) Stationary institutional functionality $U^*$ as a function of the residual activity $P$. (b) Direct, $F_{\rm dir}=\nu P S^*$ (Eq. \eqref{eq17}), and response-mediated, $F_{\rm resp}=\mu A^*S^*$ (Eq. \eqref{eq18}), contributions to the suppression of strike participation as functions of $P$. Results are shown in the resilience-paradox regime, with $\nu=0.40<\nu_c=0.96$. The remaining parameters are $\lambda=0.15, \mu=0.60, \eta=0.80, \theta=0.30$ and $\omega=0.60$. Increasing residual activity improves institutional functionality, thereby weakening the institutional response and reducing the response-mediated suppression $F_{\rm resp}$, while simultaneously strengthening the direct suppression $F_{\rm dir}$. The competition between these opposing effects constitutes the mechanism underlying the resilience paradox.}
\label{fig3}
\end{figure}

In Fig. \ref{fig3} (a), we show the stationary institutional functionality $U^*$ as a function of the residual activity $P$. The stationary functionality increases with $P$, indicating that a larger residual activity allows the institution to preserve a greater fraction of its normal operation during the strike. As a consequence, the loss of functionality becomes smaller, reducing the intensity of the stationary institutional response.

Figure \ref{fig3} (b) shows the corresponding behavior of $F_{dir}$ and $F_{resp}$. The direct suppression contribution $F_{dir}$ increases with $P$, reflecting the stronger direct effect of residual activity on strike participation. In contrast, the response-mediated contribution $F_{resp}$ decreses with $P$. This reduction occurs because larger residual activity preserves institutional functionality and consequently weakens the institutional response to the strike, as we also observed in Fig. \ref{fig3} (a). The competition between these opposing contributions constitutes the dynamical mechanism underlying the resilience paradox.

It is worth noting that the crossing of $F_{dir}(P)$ and $F_{resp}(P)$ does not define the compensation threshold. Their crossing only indicates that the two suppression contributions have equal magnitudes at a particular value of $P$. In contrast, the threshold $\nu_c$ is determined by the marginal response of the stationary strike participation to changes in $P$, as quantified by the sign of $\chi_P$. Thus, the resilience paradox is controlled not by which suppression term is larger at a given $P$, but by how the overall balance changes when residual activity is varied.

%%%%%%%%%%%%%%%%%%%%%%%%%%%%%%%%%%%%%%%%%%%%%%%%%%%%%%%%%%%%%%%%
%%%%%%%%%%%%%%%%%%%%%%%%%%%%%%%%%%%%%%%%%%%%%%%%%%%%%%%%%%%%%%%%

%%%%%%%%%%%%%%%%%%%%%%%%%%%%%%%%%%%%%%%%%%

\section{Final Remarks}

In this work, we introduced a minimal dynamical model to investigate labor strikes in partially functional institutions. The central ingredient of the model is the coexistence of a striking sector and a non-striking sector that preserves part of the institutional activity during the conflict. This residual activity affects strike participation through two competing mechanisms: a direct suppressive effect and an indirect resilience-mediated feedback associated with the reduction of institutional disruption and, consequently, of the institutional response.

The analytical treatment revealed that the competition between these mechanisms is controlled by a compensation threshold, $\nu_c = \frac{\mu\eta\omega}{\theta}$, which separates two qualitatively distinct regimes. For $\nu>\nu_c$, the conventional response is recovered, with increasing residual activity reducing stationary strike participation. In contrast, for $\nu<\nu_c$, the system enters a resilience-paradox regime, in which larger residual activity leads to an increase in stationary strike participation. This behavior was characterized through the strike-response susceptibility $\chi_P=dS^*/dP$, whose sign directly identifies the two regimes. At the compensation threshold, the direct and indirect mechanisms exactly balance and the stationary strike participation becomes independent of the residual activity $P$.

The numerical results provide a direct interpretation of this counterintuitive behavior. Increasing residual activity improves stationary institutional functionality, thereby reducing the intensity of the institutional response. As a consequence, the response-mediated suppression of strike participation decreases even though the direct suppression associated with residual activity becomes stronger. The resilience paradox therefore arises not from any paradoxical behavior of the individual agents, but from the collective balance between two individually reasonable mechanisms. In this sense, the phenomenon provides a simple example of an emergent effect in which a locally stabilizing process can produce an unexpected macroscopic consequence.

From a broader perspective, the model suggests that institutional resilience should not always be interpreted solely as the capacity to absorb disruption. In situations involving strategic collective action, the ability to preserve functionality may also modify the incentives and feedbacks that govern the persistence of the underlying conflict. By partially absorbing the effects of a strike, residual activity can reduce the pressure for a rapid institutional response and thereby contribute to the maintenance of the dispute.

The present model is intentionally minimal and therefore does not attempt to reproduce the full complexity of real labor conflicts. Residual activity is treated as an external parameter, while heterogeneity among workers, temporal changes in bargaining conditions, institutional learning, and explicit negotiation processes are neglected. These ingredients could be incorporated in future extensions, for example by allowing residual activity to evolve dynamically or by introducing feedbacks between strike participation and bargaining outcomes. Nevertheless, the simplicity of the present formulation makes it possible to identify analytically the mechanism responsible for the resilience paradox and the exact condition under which it emerges.

We therefore expect that the framework proposed here may be useful beyond the specific context of labor strikes, particularly in systems where partial functionality attenuates the immediate consequences of collective disruption while simultaneously weakening the response intended to resolve it.

% ############################################################################

\section*{Acknowledgments}

The author acknowledges partial financial support from the Brazilian scientific funding agency Conselho Nacional de Desenvolvimento Científico e Tecnológico, Brazil (CNPq, Grants 308643/2023-2 and 406820/2025-2).

\reftitle{References}

%\bibliography{bibliography}

\begin{thebibliography}{999}

%\bibitem{ref-journal}
  %Author. The title of the cited article. {\em Journal} {\bf 2008}, {\em 10}, 142--149


\bibitem{Ashenfelter1969}
O. Ashenfelter, G. E. Johnson, Bargaining Theory, Trade Unions, and Industrial Strike Activity, \textit{American Economic Association} \textbf{1969}, 59(1), 35-49. 


\bibitem{Kennan1985}  
J. Kennan, The Duration of Contract Strikes in U.S. Manufacturing, \textit{Journal of Econometrics} \textbf{1985}, 28, 5-28.


\bibitem{Kennan1986}
J. Kennan, The Economics of Strikes, \textit{Handbook of Labor Economics} \textbf{1986}, 1091-1137.

  

\bibitem{Tracy1987}
J. S. Tracy, An Empirical Test of an Asymmetric Information Model of Strikes, \textit{Journal of Labor Economics} \textbf{1987}, 5(2), 149-173.


\bibitem{Cramton1992}
P. C. Cramton, J. S. Tracy, Strikes and Holdouts in Wage Bargaining: Theory and Data, \textit{American Economic Association} \textbf{1992}, 82(1), 100-121.


\bibitem{CardOlson1995}
D. Card, C. A. Olson, Bargaining Power, Strike Duration, and Wage Outcomes: An Analysis of Strikes in the 1880s, \textit{NBER Working Paper 4075}, \textbf{1992}, $https://doi.org/10.3386/w4075$. 


\bibitem{SchnellGramm1994}
J. F. Schnell, C. L. Gramm, The Empirical Relations between Employers' Striker Replacement Strategies and Strike Duration, \textit{ILR Review}, \textbf{1994}, 47(2), 189-206.


\bibitem{Kallas2026}
J. Kallas, Deepening Our Understanding of Labor Action: Examining How Workers Organize Different Types of Strikes in the United States, \textit{Industrial Relations}, \textbf{2025}, 65, 64-79.




\bibitem{Cordova1985}
E. C\'ordova, Strikes in the public service: some determinants and trends, \textit{International labour review}, \textbf{1985}, 124(2), 163-179.


\bibitem{Rose2008}
J. B. Rose, Regulating and Resolving Public Sector Disputes in Canada, \textit{ Journal of Industrial Relations}, \textbf{2008}, 50(4), 545-559.



\bibitem{Holling1973}
C. S. Holling, Resilience and Stability of Ecological Systems, \textit{Annual Review of Ecology, Evolution, and Systematics}, \textbf{1973}, 4, 1-23.


\bibitem{LengnickHall2011}
C. A. Lengnick-Hall, T. E. Beck, M. K. Lengnick-Hall, Developing a capacity for organizational resilience through strategic human resource management, \textit{Human Resource Management Review}, \textbf{2011}, 21 243-255.



\bibitem{Castellano2009}
C. Castellano, S. Fortunato, V. Loreto, Statistical physics of social dynamics, \textit{Reviews of Modern Physics}, \textbf{2009}, 81, 591.



\bibitem{Perc2017}
M. Perc, J. J. Jordan, D. G. Rand, Z. Wang, S. Boccaletti, A. Szolnoki, Statistical physics of human cooperation, \textit{Physics Reports}, \textbf{2017}, 687, 1-51.



\bibitem{nuno1}
N. Crokidakis, The propensity for disobedience: Rule-breaking, compliance and social phase transitions, \textit{Physica A: Statistical Mechanics and its Applications} \textbf{2026}, 697, 131697.



\bibitem{nuno2}  
N. Crokidakis, When cardinals strategize: An agent-based model of influence and ideology for the papal conclave, \textit{Chaos, Solitons $\&$ Fractals}, \textbf{2026}, 209, 118411.



\bibitem{nuno3}  
N. Crokidakis, J. L. C. da C. Filho, Corruption as a self-sustained collective state in political systems, \textit{Frontiers in Physics}, \textbf{2026}, 14, 1897453.




  
\end{thebibliography}

\end{document}